\documentclass[RNAAS]{aastex631}

\usepackage{newtxtext,newtxmath}
\usepackage{amsmath}	
\usepackage{color}
\usepackage{hyperref}

\newcommand{\HLa}{H Ly$\alpha$}
\newcommand{\Lya}{Ly$\alpha$}

\begin{document}

\title{The 23.01 release of Cloudy}

\author[0000-0002-0786-7307]{Chamani M. Gunasekera}
\affiliation{Physics \& Astronomy, 
University of Kentucky, 
Lexington, Kentucky, USA}

\author[0000-0001-7490-0739]{Peter A. M. van Hoof}
\affiliation{Royal Observatory of Belgium, 
Ringlaan 3, 1180 Brussels, Belgium}


\author[0000-0002-8823-0606]{Marios Chatzikos}
\affiliation{Physics \& Astronomy,
University of Kentucky, 
Lexington, Kentucky, USA}

\author[0000-0003-4503-6333]{Gary J. Ferland}
\affiliation{Physics \& Astronomy,
University of Kentucky,
Lexington, Kentucky, USA}



\begin{abstract}
We announce the C23.01 update of {\sc cloudy}.
This corrects a simple coding error, present since $\sim 1990$, in one routine that required a conversion from the line-center to the mean normalization
of the \Lya\ optical depth. 
This affects the destruction of H~I \Lya\  by
background opacities. 
Its largest effect is upon the \Lya\  intensity
in  high-ionization dusty clouds, where the predicted
intensity is now up to three times stronger.
Other properties that depend on Ly$\alpha$ destruction, such as grain infrared emission,
change in response.

\end{abstract}



\section{Introduction} 
\label{intro}

We present the C23.01 update to {\sc cloudy}, succeeding C23.00
\citep{1998PASP..110..761F,2013RMxAA..49..137F,2017RMxAA..53..385F,2023RMxAA..59..327C}.
We correct an error in our 
use of the \citet{1980ApJ...236..609H} theory of  \Lya\  destruction that dates back to $\sim$1990.
The  \Lya\ intensity can change by as much as 0.5 dex for highly ionized dusty clouds.

\section{Line transfer in dusty media}

Resonance lines of abundant atoms often have large optical depths, so
the effects of scattering and absorption must be taken into account. 
Line photons can be lost by either collisional deexcitation, where the photon's
energy is converted into thermal energy, or by absorption by background opacities,
where the photon is absorbed by photoelectric
or grain opacity \citep{1970stat.book.....M,1992ASSL..170.....E, Rutten03, 2014tsa..book.....H}.

The escape probability formalization is a powerful treatment of the effects of radiative trapping \citep{1985ApJ...299..752N, 1992ASSL..170.....E, 2006agna.book.....O}.
This replaces the net radiative bracket 
in the radiative transfer equation with 
the product of the transition rate and
the probability that a photon will escape in a single scattering,
$\beta_{\rm esc} A_{ul}$ \citep{1985ApJ...299..752N, 1992ASSL..170.....E}. 
Here $\beta_{\rm esc}$ is the escape probability for a transition with upper and lower levels $u,l$. 
 
The line profile function $\varphi(x)$ lies at the heart of line transfer physics. 
Radiative trapping and escape compete with photon destruction through continuous absorption processes, 
the so-called ``background opacity''.
The Voigt line profile,
\begin{equation*}
    H(a,x) = \frac{a}{\pi} \int{\frac{dt e^{-t^2}}{(x-t)^2+a^2}} \, ,
\end{equation*}
describes how matter absorbs photons, while the redistribution
function describes the relation between absorption and
subsequent emission.
Here, $x$ is the frequency displacement from line-center, and $a$ is the Voigt parameter \citep{1938ApJ....88..508H, 1980ApJ...236..609H}.

There are two normalizations of the optical depth, line-center and mean.
Optical and ultraviolet spectroscopy conventionally use line-center optical depths \citep{1970stat.book.....M, Rutten03}. 
Mean optical depths are more convenient for
very large values of the damping constant, as is typical of
allowed lines of high-ionization species in the X-ray regime
\citep{1980ApJ...236..609H}. 
For small $a$, the mean optical depth is
a factor of $\sqrt\pi$ times
larger than the line-center optical depth.

{\sc cloudy} internally works with line-center optical depths throughout. 
When the code expanded to higher energies,  mean optical depths
became more useful, and mean optical depths are now reported.
{\sc cloudy} treats line overlap and integrates the
line optical depth through a medium where the Doppler width will generally be
varying, so it is not straightforward to convert the line-center optical depths
used internally to a mean optical depth. To avoid these difficulties, we
report a mean optical depth that is exactly a factor $\sqrt\pi$ times
\emph{larger} than the line-center optical depth. This assumes that the Voigt
function at line-center is $H(0, a) \approx 1$. This approximation
is valid when the damping parameter $a$ is small, $a\ll1$  
which is generally the case in the
IR, optical, and UV wavelength range. 
This conversion is more approximate
in the X-ray regime where $a$ can be much larger.

{\sc cloudy} uses fits to the escape $\beta_{\rm esc}$ and destruction 
$\beta_{\rm des}$ probabilities presented in 
\citet[hereafter HK80]{1980ApJ...236..609H} for H~I \Lya.
This theory uses mean optical depths $\tau$, and considers a wide variation in the values of $a$.  
The original incorporation of the HK80 theory into {\sc cloudy}, 
dating back to at least the C80 (circa $\sim$1990) release, 
required a conversion from $\tau$ to $\tau_0$ within one routine. 
We found that this conversion had an erroneous factor of $\pi$ in the calculation of $\beta_{\rm HK}$,
a parameter in the HK80 theory.
This affected the loss of  H~I \Lya\
photons due to ``background'' opacity, with dust being the
dominant opacity absorbing Ly$\alpha$.

In the following sections, we derive the new calculation of $\beta_{\rm HK}$, 
discuss changes in the simulated spectra, and detail how the C23.01 release can be downloaded.  

\section{Lyman \texorpdfstring{$\alpha$}{} Escape \& Destruction Probability} 
\label{derivation}

The starting point of the HK80 theory is the $\beta_{\rm HK}$ parameter, defined as (Eq.~2.7 of HK80):
\begin{equation}
    \beta_{\rm HK} \equiv k_{\rm c}/k_{\rm L},
\end{equation}
where $k_{\rm c}$ is the continuum opacity of background destruction processes 
(such as absorption on dust grains), and $k_{\rm L}$ is the line opacity. 
{\sc cloudy} uses a slightly different definition for $\beta_{\rm HK}$ 
that is better behaved in extreme conditions (like hydrogen-deficient 
and very dust-rich gas, see e.g., \citealp{Borkowski91}):
\begin{equation}
    \beta_{\rm HK} \equiv k_{\rm c}/(k_{\rm L} + k_{\rm c}).
\end{equation}
For the range of conditions considered by HK80 ($k_{\rm c} \ll k_{\rm L}$), 
there is very little difference between the two definitions. 
The opacity of the line is defined as (Eq.~2.3 of HK80):
\begin{equation}
    k_{\rm L} = \frac{N_l B_{lu} h \nu_0}{4\pi \Delta}.
    \label{kldef}
\end{equation}
Here $N_l$ is the lower-level population, $B_{lu}$ is the Einstein-B coefficient for absorption, 
$h$ is the Planck constant, $\nu_0$ is the central frequency of the line, 
and $\Delta$ is the Doppler width of the line. 
The Einstein-B coefficient is related to the oscillator strength $f_{lu}$ as follows \citep{Rybicki79}:
\begin{equation}
    B_{lu} = \frac{4\pi^2 e^2}{m_{\rm e}c \, h \nu_0} f_{lu}.
    \label{b12def}
\end{equation}
Here $e$ is the elementary charge in electrostatic units, 
$m_{\rm e}$ is the mass of the electron, and $c$ is the speed of light. 
The Doppler width is converted to velocity units $\Delta_{\rm v}$ as follows:
\begin{equation}
    \Delta = \frac{\Delta_{\rm v}}{c} \nu_0 = \Delta_{\rm v}\sigma_0,
    \label{dveldef}
\end{equation}
where $\sigma_0$ is the wavenumber of the line, $\sigma_0 = \nu_0/c$. 
Substituting Eqs.~\ref{b12def} and \ref{dveldef} into Eq.~\ref{kldef} yields:
\begin{equation}
    k_{\rm L} = N_l \frac{\pi e^2}{m_{\rm e} c \, \Delta_{\rm v} \, \sigma_0} f_{lu}.
    \label{kldef2}
\end{equation}
The opacity constant of the line $\kappa_{\rm L}$ is defined in {\sc cloudy} as:
\begin{equation}
    \kappa_{\rm L} = \frac{\sqrt{\pi} e^2}{m_{\rm e}c} f_{lu} \frac{1}{\sigma_0}.
    \label{kappadef}
\end{equation}
Substituting Eq.~\ref{kappadef} into Eq.~\ref{kldef2} yields:
\begin{equation}
    k_{\rm L} = N_l \kappa_{\rm L} \sqrt{\pi} / \Delta_{\rm v}.
    \label{kldef3}
\end{equation}
In the previous {\sc cloudy} code, $\beta_{\rm HK}$ was calculated using the $k_L$ 
derived in Equation~\ref{kldef3} divided by $\pi$. 

\section{Changes to Spectra}
\label{results}

With this change, $k_L$ is now larger by a factor of $\pi$, 
decreasing $\beta_{\rm HK}$ by approximately the same factor 
(assuming $k_{\rm c} \ll k_{\rm L}$). 
Thus, the destruction probability by background opacities is now reduced by approximately the same factor of $\pi$.

\begin{figure}
\centering
    \includegraphics[width=0.65\columnwidth]{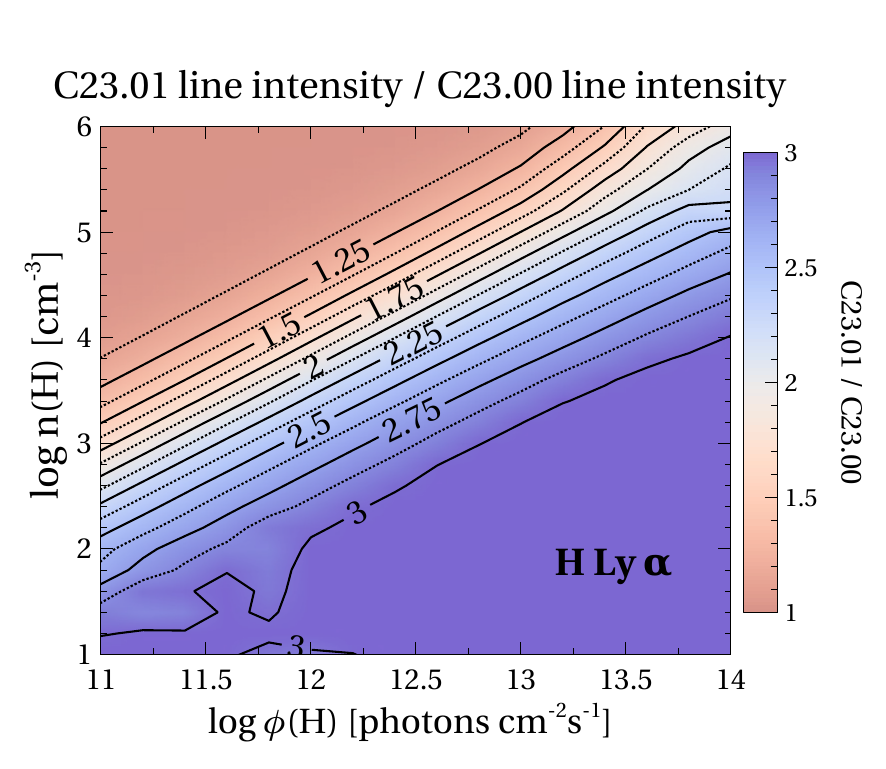}
    \caption{A contour plot of \HLa\ line intensity predicted by C23.01 relative to the same quantity from C23.00, 
    for the baseline model \texttt{orion\_hii\_open.in} in the {\sc cloudy} test suite. 
    The ratio of ionizing photon flux $\phi(H)$ to hydrogen density $n(H)$ is effectively the ionization parameter $U$. The lower-right corner of the panels corresponds to high $U$, and the upper-left corner is low $U$. }
    \label{contour}
\end{figure}

{\sc cloudy} includes a large test suite which makes autonomous testing possible. These showed that the largest changes occurred for dusty clouds where \HLa\ destruction is important. Figure~\ref{contour} shows the \HLa\ intensity for a blister H~II region model 
inspired by the Orion H II region, 
a hydrostatic ionized layer on the surface of a molecular cloud
\citep{1991ApJ...374..580B}. 
We consider a range of hydrogen densities ($10\leq n({\rm H}) \leq 10^6$ cm$^{-3}$) and stellar ionizing photon fluxes ($10^{11}\leq \phi({\rm H}) \leq 10^{14}$ photons\,cm$^{-2}$\,s$^{-1}$). 
The Figure shows predictions of C23.01 
relative to those of C23.00. 

Our new calculations have reduced the destruction probability
so \HLa\ is now stronger.
The biggest change occurs at high $\phi(H)$ and low $n(H)$, corresponding to high ionization parameters ($U=\phi({\rm H}) / c n({\rm H})$). 
As we increase $U$ and the ionization, grains proportionately absorb more of the ionizing photons 
\citep{1998PASP..110.1040B, 2023MNRAS.520.4345G}.
More highly ionized clouds have less atomic opacity but the same grain opacity, so there is a
greater probability of a photon being absorbed by grains rather than scattered from a hydrogen atom.
The destruction probability of \HLa\ is higher at high $U$.
Reducing the destruction probability by a factor of $\pi$ results in the greatest enhancement of \HLa\ in the lower right corner.  

\section{Summary}
\label{summary}

The primary result of this update are changes to the \HLa\ line in dusty environments at moderate to high ionization parameter. 
We present an update to {\sc cloudy}, which includes changes in the hydrogen Lyman $\alpha$ escape and destruction probabilities based on HK80. 
Since \HLa\ is often the strongest line in the spectrum, significant changes in its physics will have consequences on other physical parameters.
A discussion of the changes to the grain emission and other consequences will be presented in the C24 release paper.

This update has been released as C23.01 and is now available to download from \href{https://gitlab.nublado.org/cloudy/cloudy/-/wikis/home}{wiki.nublado.org}. Papers that use this version of {\sc cloudy} should cite both \citet{2023RMxAA..59..327C} and the present paper.

\begin{acknowledgments}
CMG acknowledges support by NASA (19-ATP19-0188).
MC acknowledges support from NSF (1910687), and NASA (19-ATP19-0188, 22-ADAP22-0139).
GJF acknowledges support by NSF (1816537, 1910687), NASA (ATP 17-ATP17-0141, 19-ATP19-0188), and STScI (HST-AR-15018 and HST-GO-16196.003-A).
\end{acknowledgments} 

%

\vspace{5mm}







\bibliography{sqrtpi_bug_fix2}{}
\bibliographystyle{aasjournal}



\end{document}